\documentstyle[12pt]{article}

\advance\textheight by \topskip
\newcommand{\Dir}{\kern -6.4pt\Big{/}}
\newcommand{\Dirin}{\kern -10.4pt\Big{/}\kern 4.4pt}
\newcommand{\DDir}{\kern -7.6pt\Big{/}}
\newcommand{\DGir}{\kern -6.0pt\Big{/}}
\newcommand{\ra}{\rightarrow}
\newcommand{\be}{\begin{equation}}
\newcommand{\ee}{\end{equation}}
\newcommand{\beq}{\begin{eqnarray}}
\newcommand{\eeq}{\end{eqnarray}}

\begin{document}

\thispagestyle{empty}
\setcounter{page}{0}

\begin{flushright}
{\large DFTT 49/95}\\
{\rm August 1995\hspace*{.5 truecm}}\\
{\large hep-ph/9509282}\\
\end{flushright}

\vspace*{\fill}

\begin{center}
{\Large \bf NLO QCD corrections to the production of off--shell WW pairs at
$e^+ e^-$ colliders \footnote{ Work supported in part by Ministero
dell' Universit\`a e della Ricerca Scientifica. \hfill\break\indent
\ e--mail: maina@to.infn.it, pizzio@to.infn.it}}\\[2cm]
{\large Ezio Maina and Marco Pizzio}\\[.3 cm]
{\it Dipartimento di Fisica Teorica, Universit\`a di Torino}\\
{\it and INFN, Sezione di Torino}\\
{\it v. Giuria 1, 10125 Torino, Italy.}\\
\end{center}

%

\vspace*{\fill}

\begin{abstract}
{\normalsize
QCD corrections to
$e^+e^- \ra WW \ra q_1 \bar q_1 q_2 \bar q_2$ are computed
and presented in a form which allows to impose
realistic cuts on the structure of the observed events.
QCD radiation substantially modifies the jet--jet invariant
mass distributions from which the value of the $W$ mass will be extracted.
When the range of allowed jet--jet masses is restricted, as it has been
proposed
in order to suppress non--$WW$ backgrounds, the total cross section is also
deeply affected. If all events are forced to four jets, combining the
two partons with smallest invariant mass, and the reconstructed masses are
required to satisfy $ \left| M_{Ri} -M_W \right| \le 10$ GeV $i$=1,2 the
lowest order cross section can be reduced by more than 40\%.
}
\end{abstract}

\vspace*{\fill}

\newpage
\subsection*{Introduction}
\par
One of the main goals of Lep II is the measurement of the W mass
with high accuracy, possibly of the order of 50 MeV or less
\footnote{For a general introduction to the Standard Model predictions for
$W$--pair production in $e^+e^-$ collisions see \cite{bd}.}.
Two methods have been singled out as the most promising \cite{old_exp}.
The first one is based on the rapid increase of the total cross section at
threshold and has been recently studied in detail in
ref. \cite{Stirling_threshold} where it has been shown that the optimal
collider energy is about 161 GeV. The second method relies on the direct
reconstruction of the mass from the hadronic decay products of the $W$ using
the decay channels
\beq
W^+ W^- &\ra& q_1 \bar q_1 q_2 \bar q_2 \\
W^+ W^- &\ra& q \overline{q} \ell \nu
\eeq
where $\ell = e, \mu$. The tree--level
relevant branching ratios are given in table I.
\par
The threshold method will be the first one to be exploited
when Lep II is turned on. In the meantime
the number of superconducting cavities will be increased until the
center--of--mass energy can be pushed to about 175 GeV where the
much larger cross--section will allow $M_W$ to be extracted by direct
reconstruction.
Preliminary studies indicate that the direct measurement
will provide a more precise determination of the $W$ mass than the
threshold method if all decay channels are combined, while the two methods
are expected to have comparable accuracies if only the semileptonic decays
are used for direct reconstruction.
The possibility of a further energy upgrade to about 190
GeV is presently under consideration. This higher energy would greatly improve
the prospects for the discovery of new particles, in particular of the Higgs
boson, without compromising the precision expected for the measurement of
$M_W$.
\par
In the case of semileptonic decays, apart from all the usual difficulties
related to measuring jet energies and directions, the main uncertainty in
the mass measurement derives from the reconstruction of the unobserved
neutrino. These uncertainties can be
partially eliminated by using energy--momentum conservation and possibly
the approximate equality of the $W^+$ and $W^-$ masses.
In the case of hadronic decays, where in principle all decay products can be
detected, there are additional ambiguities which
stem from the fact that two decays occur in the same event. Even in the
simplest
case, in which only four jets are detected, there are three possible ways
of pairing the jets. If more jets are present the combinatorics becomes rapidly
quite complicated.
\par
It has been proposed to force all events to four jets simply
iterating a given reconstruction algorithm until only four clusters are left.
It is obvious that such a scheme is not without perils.
A jet from, say , the decay of the $W^+$ can be closer
to a jet from the decay of the $W^-$ than to
any other jet from the positively charged $W$. If the stray jet carries
large energy, then the reconstructed masses will be quite far from the
$W$ mass which is already at present known with an error smaller than 200 MeV
\cite{W_mass}.
One could imagine that such events could be discarded, at the same time
providing a mean of suppressing non--$WW$ background.
However, due to the intrinsic width of the $W$ and to all
experimental uncertainties, it is impossible to impose very stringent
cuts on the difference between the measured jet--jet masses and the true mass
without substantially reducing the event rate.
A precise estimate of the effect of these mass cuts is necessary in order
to asses the final accuracy with which $M_W$ will be measured.
\par
A different proposal has been to simply discard all events with five or more
jets. It is clear that this can only be a partial solution at best.
It excludes hard, non--collinear gluon emission but it does not suppress
soft and/or collinear radiation whose effects need in any case to be studied.
Furthermore such a procedure would introduce a dependence of the mass
measurement on the invariant mass cut which separates four--jet events
from the events with a larger jet multiplicity.
\par
QCD radiation can modify the characteristics of the final state like
the distributions of jet momenta, jet--jet invariant masses and opening
angles from which the value of the $W$ mass will be extracted.
These semi--inclusive quantities, on the other hand, are of calorimetric type,
that is infrared and collinear safe. The measured invariant masses, as an
example, do not change if a massless parton of momentum $p$ splits into two
partons of momentum $\lambda p$ and $(1-\lambda ) p$
or if an additional soft gluon is radiated by any of the hard partons.
As a consequence they can be reliably studied in perturbative QCD.
It is well known that differential distributions, particularly when the the
phase space for gluon emission is restricted by experimental cuts, can be more
sensitive to higher order corrections than fully inclusive quantities like
total
cross--sections in which virtual and real contributions cancel to a large
degree. It is therefore necessary to include higher order QCD effects into the
predictions for $WW$ production and decay in a way which allows to impose
realistic cuts on the structure of the observed events.
In this paper we present the calculation of the complete $\cal{O}(\alpha_s)$
corrections to the three diagrams which describe the basic process
$e^+e^- \ra WW \ra q_1 \bar q_1 q_2 \bar q_2$.
We will discuss only fully hadronic
decays but the extension to semileptonic decays is trivial.
\par
Recently it has been pointed out \cite{gpz,sjo_kho,gu_ha,bose}
that cross--talk between the decay of the two $W$'s
might take place also at energy scales much smaller than those typical
of jets through color reconnection phenomena and Bose--Einstein correlations.
This could weaken the notion of two separate decays and cast doubts on
our ability to reconstruct the masses of the original sources from the decay
products, producing potentially large uncertainties.  Our understanding of
these
issues is, at present, extremely poor, and the predictions are strongly
model--dependent. In this letter we will leave aside these non--perturbative
issues. For a discussion of some aspect of colour reconnection from a
perturbative point of view see \cite{6j,ww_colour}.

\subsection*{Calculation}
In $n$ dimensions, $n = 4 - 2\epsilon$, using dimensional regularization,
the tree--level
amplitude squared for the decay $W \ra q \bar q $ is:
\be\label{Aqq}
\left| M_0 (q \bar q ) \right|^2 = 4 K s (1 - \epsilon)
\ee
with $K = N_C e^2 e_q^2$ and, labelling the momenta with the particle names,
$s= (q + \bar q)^2$.
The corresponding amplitude squared for the decay $W \ra q \bar q g$ is:
\be\label{Aqqg}
\left| M (q \bar q g) \right|^2 =
8 K^\prime (1 - \epsilon)\left\{
 \left( \frac{2 s^2}{s_1 s_2} -\frac{2 s}{s_1} -\frac{2 s}{s_2}
 + \frac{s_2}{s_1} + \frac{s_1}{s_2} \right)
- \epsilon \left(\frac{s_2}{s_1} + \frac{s_1}{s_2} + 2 \right)
\right\}
\ee
with $K^\prime = K C_F  g_s^2 \mu^{2\epsilon}$, $s= (q + \bar q + g)^2$,
$s_1= (q + g)^2$ and $s_2= ( \bar q + g)^2$.
\par
The phase space can be expressed as follows:
\beq\label{ph_space}
\hspace{-.2cm} d \makebox{{\rm Lips}} (Q \ra q + \bar q + g) &=&
 d \makebox{{\rm Lips}} (Q \ra q + \bar q ) \times \nonumber \\
& & \frac{s^{3\epsilon -1}}{16 \pi^2}
\frac{(4 \pi)^\epsilon}{\Gamma (1 - \epsilon)}
 d s_1  d s_2
\left\{(s-s_1-s_2)s s_1 s_2 \right\}^{-\epsilon}
\eeq
after integrating over the azimuthal angle in the center of mass of
$q$ and $g$ with $\bar q$ on the $z$--axis.
\par
Integrating over $s_1$ and $s_2$ in the region $\Sigma$ where
$\min ( s_1,s_2) < \Delta$ one obtains:

\beq\label{ph_space_int}
& & \hspace{-.5cm}
\left| M_{\Sigma} (q \bar q ) \right|^2 =
\frac{s^{3\epsilon -1}}{16 \pi^2}
\frac{(4 \pi)^\epsilon}{\Gamma (1 - \epsilon)}
\int\!\!\int_{\Sigma}  d s_1  d s_2
\left\{(s-s_1-s_2)s s_1 s_2 \right\}^{-\epsilon}
\left| M (q \bar q g) \right|^2
\\
& & \hspace{-.5cm}
=\left| M_0 (q \bar q ) \right|^2 C_F \frac{g_s^2}{4\pi^2} \xi
\left\{ \frac{1}{\epsilon^2} + \frac{3}{2\epsilon} - \ln^2 \eta
+ \ln\left( \frac{\eta}{1-\eta}\right) \left(2\eta -\frac{1}{2}\eta^2
-\frac{3}{2} \right) \right. \\
& & \hspace{3.5cm}
\left. +\frac{7}{2} +\frac{5}{4}\eta -2 \,\makebox{{\rm Li}}_2 (\eta )
-\frac{5}{2} \,\makebox{{\rm Li}}_2 (1)
\right\} \nonumber \\
\label{ph_space_int_approx} & & \hspace{-.5cm}
=\left| M_0 (q \bar q ) \right|^2 C_F \frac{g_s^2}{4\pi^2} \xi
\left\{ \frac{1}{\epsilon^2} + \frac{3}{2\epsilon} - \ln^2 \eta
- \frac{3}{2}\ln \eta + \frac{7}{2} - \frac{5}{2}
\,\makebox{{\rm Li}}_2 (1)
\right\}  + {\cal O} (\eta\ln\eta )
\eeq
where $\eta= \Delta / s$ and $\xi= \left(\frac{\mu^2}{s}\right)^\epsilon
e^{-\epsilon\left(\gamma-\ln (4\pi) \right)}$.
$\makebox{{\rm Li}}_2 (x)$ is the standard dilogarithm with
$\makebox{{\rm Li}}_2 (1) = \pi^2/6$. $\Delta$ separates the region of soft
and collinear emission from the region where gluon radiation is
considered hard. The dependence of the result on $\eta$ will be discussed
later in detail.
\par
The $\cal{O}(\alpha_s)$ one--loop contribution to the $W$ decay is
\be\label{virtual}
M_V (q \bar q ) = M_0 (q \bar q )
C_F \frac{g_s^2}{8 \pi^2} \xi
\left\{ -\frac{1}{\epsilon^2} - \frac{3}{2\epsilon} - 4
      + \frac{7}{2}\, \makebox{{\rm Li}}_2 (1)
- i \pi \left( \frac{1}{\epsilon} +\frac{3}{2}\right)
\right\}.
\ee
Combining the virtual contribution (\ref{virtual}) with the integral of the
real--emission cross section over the soft and collinear region
(\ref{ph_space_int}--\ref{ph_space_int_approx}) the following simple
expression is obtained for the $W$ decay at $\cal{O}(\alpha_s)$:
\beq\label{W-decay}
\left| M (W \ra q \bar q ) \right|^2  &=& \left| M_0 (q \bar q ) \right|^2
+ 2 \makebox{{\rm Re}}\left( M_0^\ast (q \bar q ) M_V (q \bar q )\right)
          + \left| M_{\Sigma} (q \bar q ) \right|^2
          + \left| M_{\overline \Sigma} (q \bar q ) \right|^2  \\
&=&
    \left| M_0 (q \bar q ) \right|^2 \left( 1 + F(\eta) \right)
          + \left| M_{\overline \Sigma} (q \bar q ) \right|^2  \\
&=&
   \left| M_0 (q \bar q ) \right|^2
          \left( 1 + F^\prime (\eta) \right)
          + \left| M_{\overline \Sigma} (q \bar q ) \right|^2
+ {\cal O} (\eta\ln\eta )
\eeq
where
\beq \label{F}
F(\eta ) &=& \frac{C_F \alpha_s }{\pi}
\left\{ - \ln^2 \eta
+ \ln\left( \frac{\eta}{1-\eta}\right) \left(2\eta -\frac{1}{2}\eta^2
-\frac{3}{2} \right) \right. \\
& &
\left. \hspace{5.5cm} -\frac{1}{2} +\frac{5}{4}\eta
-2 \,\makebox{{\rm Li}}_2 (\eta )
+\makebox{{\rm Li}}_2 (1)
\right\}, \nonumber \\
\label{F_prime}
F^\prime (\eta ) &=& \frac{C_F \alpha_s }{\pi} \left(
- \ln^2 \eta - \frac{3}{2}\ln \eta - \frac{1}{2} + \makebox{{\rm Li}}_2 (1)
              \right).
\eeq
In the previous formulae
 $\left| M_{\overline \Sigma}(q \bar q )\right|^2$ is defined
in analogy to (\ref{ph_space_int}) as the integral of the
real--emission amplitude squared $\left| M (q \bar q g) \right|^2$
over the hard--gluon region $\overline\Sigma$
where $\min ( s_1,s_2) > \Delta$. The expression for the sum of the virtual
and soft--collinear contributions has been derived many times before in
various contests (see for example \cite{Glover_Stirling,Kramer_Lampe})
and we are in agreement with previous results, possibly after
a simple coupling redefinition.\par
Exploiting the universality of soft and collinear divergencies
\footnote{The formalism is discussed in detail in ref. \cite{Giele_Glover}.},
these results can be directly carried over to $e^+e^- \ra WW \ra
q_1 \bar q_1 q_2 \bar q_2$. At $\cal{O}(\alpha_s)$ there is no
interference between the decays of the two $W$'s and the amplitude for
$e^+e^- \ra WW \ra q_1 \bar q_1 q_2 \bar q_2 g$
splits into two orthogonal terms $M_1$ and $M_2$
describing the emission
from the $q_1 \bar q_1$ pair and the $q_2 \bar q_2$ pair
respectively.
The quantity $s$ in (\ref{W-decay}) in the two terms has to be identified with
the invariant mass $s_{1g}$ of the $q_1 \bar q_1 g$ system and
the invariant mass $s_{2g}$ of the $q_2 \bar q_2 g$ system respectively.
Choosing
$\eta_1= \Delta_1 / s_{1g} = \eta_2= \Delta_2 / s_{2g} = \eta$ we can write:
\beq \label{WW-4q}
\left| M (WW \ra 4q ) \right|^2 &=& \left| M_0 \right|^2
+ 2 \makebox{{\rm Re}}\left( M_0^\ast  M_V \right)
          + \left| M_{\Sigma}  \right|^2
          + \left| M_{\overline \Sigma}  \right|^2 \\
\label{WW-4q-approx}
&=& \left| M_0 \right|^2
          \left( 1 +2 F^\prime (\eta ) \right)
          + \left| M_{\overline \Sigma}\right|^2
+ {\cal O} (\eta\ln\eta )
\eeq
In (\ref{WW-4q}) $M_0$ is the tree-level amplitude and
$M_V$ the one--loop  $\cal{O}(\alpha_s)$ amplitude for $e^+e^- \ra WW \ra 4q$.
$\left| M_{\Sigma}  \right|^2$ and
$\left| M_{\overline \Sigma}  \right|^2$ are obtained integrating the
tree--level amplitude squared describing $e^+e^- \ra WW \ra
q_1 \bar q_1 q_2 \bar q_2 g$ over the gluon variables as in
(\ref{ph_space_int}). $\left| M_{\Sigma}  \right|^2$ is equal
to the sum of the integral of $\left| M_1 \right|^2$ over the region $\Sigma_1$
where $\min ((q_1+g)^2,(\bar q_1+g)^2) < \Delta_1$ and of the integral of
$\left| M_2 \right|^2$ over the region $\Sigma_2$
where $\min ((q_2+g)^2,(\bar q_2 +g)^2) < \Delta_2$.
$\left| M_{\overline \Sigma}\right|^2$ is equal to the integrals of the square
of $M_1$ and $M_2$ over the complementary regions.
These latter integrals and the tree--level matrix element
need to be calculated only
in four dimensions.
\par
The real emission matrix element squared $\left| M(X+g)\right|^2$
factorizes into the product of the corresponding emissionless expression
$\left| M(X)\right|^2$ times a universal function which describe the
radiation of a gluon only in the soft and collinear limit.
Therefore in the expression for the sum
of the virtual and soft--collinear corrections
there are unavoidable inaccuracies of order $\eta$,
related to the approximation of neglecting the gluon energy
or emission angle
inside $\left| M(X)\right|^2$, and the use of the approximate
expression (\ref{WW-4q-approx}) produces no loss in precision.
In addition it must be remembered that we are effectively
lumping the soft--collinear region in the phase space point corresponding to
the absence of gluon radiation. Therefore, in order to obtain
an accurate determination of the total cross section and
in order to avoid disrupting
the shape of the distributions of the different observable quantities
a small value of $\eta$ must be chosen.
In all our figures we have used $\eta = 1. \times 10^{-4}$ and we have
checked that our result depend extremely weakly on the value of $\eta$ in
the range $ 5. \times 10^{-5} \le \eta \le  1. \times 10^{-3}$.
\par
Taking advantage of the fact that close to the appropriate
soft--collinear regions, which contribute most to the result,
the singular part of the integrands behaves like
$\left| M (q \bar q g) \right|^2$ (\ref{Aqqg}) we can use
$y_{1,i} = \ln (q_i+g)^2$ and $y_{2,i} = \ln ( \bar q_i+g)^2$ as
integration variables. In this way it becomes possible to evaluate
the real emission cross section using a reasonable amount of CPU time,
producing theoretical predictions of an accuracy adequate to the expected
precision of the measurement of the $W$ mass.
\par
In the case of real emission, in order to extract two candidate masses from
the event, the two partons with smallest \cite{jade}
\be\label{y_jade}
y= \frac{2 E_i E_j \left( 1 - \cos \theta_{ij} \right)}{s}
\ee
are merged, summing the corresponding four--momenta.
This corresponds, at the present order in
perturbation theory, to forcing all events to four clusters before attempting
to reconstruct the masses of the two $W$ bosons. In this paper we have
restricted our attention to the simple JADE scheme and it remains to be
studied whether a different reconstruction scheme \cite{durham,geneva}
might prove more appropriate.
Finally, the reconstructed masses
are determined, out of the three possible combinations,
choosing the two two--jet pairs whose masses
$M_{R1}$  and $M_{R2}$ minimize
\be\label{selection}
\Delta_M = \left| M_{R1} - M_W \right| + \left| M_{R2} -M_W \right|.
\ee
It is obvious that the precise form of the metric which defines the distance
in jet--space is to some extent arbitrary. For instance one could use
\be\label{selection2}
\Delta^\prime_M = \left( M_{R1} - M_W \right)^2 + \left( M_{R2} -M_W \right)^2.
\ee
We have checked that at $\sqrt{s} = 160$ GeV and at $\sqrt{s} = 175$ GeV
the cross sections obtained with (\ref{selection}) and those obtained with
(\ref{selection2}) differ by less than 1\%.
\par
The expressions presented in this section can be applied without modification
to all four--quark neutral current processes
$e^+e^- \ra ZZ, Z \gamma, \gamma\gamma
\ra q_1 \bar q_1 q_2 \bar q_2$.
\par
In order to expose the effect of ${\cal O} (\alpha_s)$ corrections we have
made a number of simplifying assumptions. (1) Initial--state--radiation (ISR)
effects have been neglected. It would however be straightforward to include
them using standard techniques \cite{ISR}.
(2) Coulomb corrections have not been
included although, since they can be expressed as a
multiplicative factor times the lowest order (LO) matrix element squared,
their inclusion would be rather simple \cite{Coulomb}.
(3) Quark masses have been neglected since the contribution of $b$--quarks
is severely suppressed by the smallness of the $V_{bc}$ element of
the CKM matrix and the $c$--quark mass is so small compared with $M_W$.
(4) We have not taken into account non--resonant electroweak contributions
and QCD four--jet backgrounds \cite{Bardin_bkg,excalibur,Stirling_threshold}.
The former has been shown
to be well below the per--cent level\footnote{However they are intimately
related to the gauge invariance of the theory. For a detailed discussion of
this issue see \cite{bd,ACO,gauge}.}
(with the possible exception of final
states including electrons or electron--neutrinos). The latter, which
is dominated by $q \bar q gg$ production, is strongly suppressed when
two jet pairs with masses close to $M_W$ are required. In
ref. \cite{Stirling_threshold} the QCD background, with $y_{cut} = 0.01$ and
requiring two jet--jet masses within 10 GeV of $M_W$, has been
estimated at about 0.1 $pb$ for $\sqrt{s} = 161$ GeV. Since the QCD
background scales only as $s^{-1}$ this estimate can be considered valid,
as a first approximation, in the full range of operation of Lep II.
However this procedure for reducing the four--jet background has far reaching
consequences on the $WW$ signal as will be discussed at length in the
following.
(5) Non--perturbative colour--reconnection and Bose--Einstein
effects have not been considered.
(6) We have ignored all experimental uncertainties in the
reconstruction of jet momenta and directions which will have to be studied
with a full detector simulation.\par
All tree--level matrix elements have been computed, in the unitary gauge,
using the formalism
presented in ref. \cite{method} with the help of a set of routines,
called PHACT \cite{phact}, which generate the building blocks of the helicity
amplitudes semi--automatically.
The matrix element for $e^+e^- \ra WW \ra
q_1 \bar q_1 q_2 \bar q_2 g$ has been first computed in ref. \cite{Brown}.
Our result has been checked against the corresponding
amplitude generated by Madgraph \cite{Madgraph}.
\par
The numerical values of the Standard Model parameters used in the calculation
are given in Table II.
\par

\subsection*{Results}
Our results are presented in fig.1, fig.2, Table III and Table IV.
The total hadronic cross sections at tree--level and
at next--to--leading--order (NLO)
in QCD are given in the first column of Table III for $\sqrt{s} = 160$ GeV
and 175 GeV.
These results are sensitive to events with jet pairs of
arbitrarily large and small invariant masses within the kinematical limits.
In practice events with jet--jet masses too far from the $W$ mass would
not be accepted as $WW$ events. Cuts of this sort have been suggested
in order to eliminate most of the four--jet QCD background.
We therefore require that each of the reconstructed
masses, selected according to (\ref{selection}), satisfies
\be\label{mass_cut}
\left| M_{Ri} -M_W \right| \le \delta, \hspace{1cm} i=1,2.
\ee
In the second and third column of Table III we show the cross section
obtained with $\delta = 30$ GeV and $\delta = 10$ GeV respectively.
In parentheses we give the cross sections obtained when all events which are
identified as containing five jets, with $y_{cut} = 1.\times 10^{-2}$ in
the JADE scheme, are discarded. The ratio of the corresponding
five--jet cross section to the total hadronic cross section
\be
R_5 = \frac{\sigma(WW\ra 5j)}{\sigma(WW\ra had)}
\ee
is about 22\% at $\sqrt{s} = 175$ GeV with no mass cut.
\par
Let us mention that in the so called ADLO--TH
set of cuts, the semi--realistic set which the Working Groups on
LEP II Physics have agreed on in order to provide a common ground for the
comparison between theoretical calculations and the simulations performed
by the experiments, two jets are considered as distinguishable if the
invariant mass of the pair is larger than 5 GeV. This corresponds to
$y_{cut} \approx 1.\times 10^{-3}$. However at this value of $y_{cut}$ the
five--jet cross section is larger than the total hadronic cross section and,
as a consequence, the cross section for events with at most four jets is
negative. This can be readily seen from eq. (\ref{F_prime}). If we neglect
for simplicity the interplay between the two decays, we have that
$y_{cut} = 1.\times 10^{-3}$, since $\sqrt{s} \approx 2 M_W$, corresponds to
$\eta\approx 4.\times 10^{-3}$ and one finds that, with $\alpha_s = .117$,
$F^\prime (0.004) < - 1$.
\par
In the narrow width, or stable $W$'s, approximation the QCD correction to the
total cross section is simply twice the correction to the $W$ decay, namely
the Born cross section gets multiplied by $(1+2\alpha_s / \pi)$.
In principle,  for off--mass--shell $W$'s, one could expect additional terms
of order $(\alpha_s \Gamma_W) / (\pi  M_W ) \approx 1.\times 10^{-3}$.
In order to study effects of this order of magnitude, all
matrix elements have been integrated using VEGAS with a relative
precision not worse than $1.\times 10^{-4}$.
We have found that the corrections to the total hadronic cross section for
unstable $W$'s do
not differ from those for stable $W$'s by more than one part in a thousand.
Therefore the narrow width approximation appears to be perfectly adequate
for the QCD correction to the total cross section \footnote{As a consequence,
the QCD corrected cross section for $W^+ W^- \ra q \bar{q} \ell \nu $,
if no restriction is imposed
on the hadronic part of the final state, is $\sigma_0(1+\alpha_s/ \pi)$
where $\sigma_0$ is the Born result.}.
This result however relies on the delicate cancellation between
real and virtual contributions, and is strongly modified by any
constraint which decreases the phase space available for real emission.
This is evident from the second and third column of Table III.
At $\sqrt{s} = 175$ GeV , with $\delta = 10 $ GeV, NLO QCD corrections
decrease the corresponding tree--level cross section by about 33\%,
which is to be contrasted with the increase of about 7\% they produce
in the total hadronic cross section in column one. Even with the much milder
cut $\delta = 30 $ GeV the LO result is decreased by about 11\%.
At $\sqrt{s} = 160$ GeV, about 200 MeV above the nominal threshold with the
value of $M_W$ adopted in this paper, the NLO result with $\delta = 10\ (30)$
GeV is 45\% (15\%) smaller than the LO prediction without mass cuts.
The corresponding cross section is decreased by approximately
0.8 (0.3) $pb$. It should be remembered that a change in the total
cross section of 0.2 $pb$ corresponds to a shift of about 100 MeV in the
measured $W$ mass \cite{Stirling_threshold}. Therefore QCD corrections,
in contradiction to the naive expectations based on the
behaviour of the total cross section, are potentially as large as the
corrections due to ISR and much larger than those produced by the Coulomb
interaction,
even though they only affect about half of the total $e^+e^-\ra W^+W^-$
cross section. Obviously, any modification of the LO cross section which can
be reliably computed can be incorporated into the theoretical predictions
which are fitted to the data and do not degrade the expected accuracy
of the measurement of $M_W$. They however directly affect the statistical
error and the uncertainty in the theoretical analysis has to be included
in the evaluation of the systematic error.
\par
The behaviour of the LO and NLO hadronic cross section as a function
of the center--of--mass energy can be found in fig.1.
As in Table III we present the total hadronic cross sections at tree--level
(dashed curve) and at NLO (continuous curve).
The long--dash--short--dash curve and the dash--dot curve
show the NLO cross sections obtained with $\delta = 30$ GeV and
$\delta = 10$ GeV respectively.
The dotted line gives the tree--level cross section with
$\delta = 10$ GeV. The analogous cross section with  $\delta = 30$ GeV,
as expected from the fact that in this case $\delta \gg \Gamma_W$,
differs very little from the total cross section and we omit it.
\par
The effect of the mass cut (\ref{mass_cut}) reaches a maximum
at about $\sqrt{s} = 155$ GeV and decreases with increasing
center--of--mass energy. We notice that at $\sqrt{s} = 150$ GeV,
with $\delta = 10$ GeV, the NLO result becomes larger than the LO
one. This is due to the fact that typically, below threshold,
one of the $W$'s is on mass--shell while the other is highly
virtual. Therefore at tree--level most of the events fail the cut.
In this case the rearrangement of invariant masses introduced by QCD
radiation blurs the separation between on--mass--shell and
off--mass--shell $W$'s and reduces the effectiveness of the mass cut.
{}From fig.1 we read that the
NLO order result with $\delta = 30$ GeV is always considerably
smaller the tree--level
prediction, confirming that even moderate restrictions on the structure of
the final state can produce sizable effects. Choosing $\delta = 10$ GeV
decreases the available event rate by a factor close to 50\% and might
result in a substantial increase in the statistical error of the $W$--mass
measurement. Just above threshold the QCD corrections depend only weakly
on the center--of--mass energy. At $\sqrt{s} = 160$ GeV they amount to
45\% (15\%) with $\delta = 10\ (30)$ GeV while at $\sqrt{s} = 165$
they become 38\% (14\%). Therefore we do not expect the determination
of the optimal energy for the threshold measurement of $M_W$
of ref. \cite{Stirling_threshold} to be modified by QCD effects.
\par
In fig.2 we present the distribution of the average of the two
reconstructed $W$ masses
\be\label{average_M}
\overline M = \frac{1}{2} \left( M_{R1} + M_{R2} \right).
\ee
which has been extensively used as an estimator of $M_W$.
At tree--level the  difference between $\overline M$ and the average
of the masses of the two virtual $W$'s, namely the masses which one would
reconstruct if the quarks could always be correctly paired,
is extremely small. Clearly one could consider more sophisticated
approaches, for instance using the mass distribution in the
$\left(M_{R1},  M_{R2}\right)$ plane but this analysis is beyond
the scope of this letter.
The two upper curves in fig.2a are obtained when all events are retained,
even when one or
both the candidate masses are far from $M_W$. The results for
$\delta = 30$ GeV and $\delta = 10$ GeV are shown in fig.2b and fig.2c
respectively. The dashed lines are the LO result while the NLO
predictions are given by the continuous lines. The NLO cross section for events
with at most four jets, at $y_{cut} = 1. \times 10^{-2}$ is shown
by the dotted lines.
When no cut on $\left| M_{Ri} -M_W \right|$ is imposed,
at NLO the peak cross section is strongly reduced with respect to
the LO distribution and a large tail at low
masses is generated.
As $\delta$ decreases the tail gradually disappears while the number of events
contained in the peak at about the $W$ mass decreases only slightly. It is
precisely the distortion of the average--mass tree--level spectrum that
explains the dramatic effect of the mass cut (\ref{mass_cut}) at NLO.
We see that the low--mass tail persists also in the lower multiplicity sample
and therefore it is not predominantly populated by five--jet events.
\par
In an attempt to quantify the modifications of the average mass
distribution which are induced by QCD corrections
we present in Table IV the average deviation,
$\langle\overline M - M_W\rangle$ and the
root--mean--square (RMS) deviation,
$\langle\left(\overline M - M_W\right)^2\rangle^{\frac{1}{2}}$,
of the mass distribution with respect to $M_W$ in GeV.
In each column the left (right) hand side value refers to the LO (NLO).
In the first row all events are accepted while in the second and third
row the two jet--jet masses satisfy
$ \left| M_{Ri} -M_W \right| \le \delta, i=1,2 $ with
$\delta = 30$ GeV and $\delta = 10$ GeV respectively. The results
in the last three rows are obtained discarding
all events which are recognized as
five jets with $y_{cut} = 1.\times 10^{-2}$ in the JADE scheme.
\par
Table IV shows that already at tree--level the average reconstructed mass
$\overline M$ is smaller than $M_W$.
The difference $\Delta M = M_W - \overline M$ depends
on the mass cut. $\Delta M$ can be larger
than 300 MeV and decreases with $\delta$
down to about 40 MeV at $\delta =  10$ GeV.
This behaviour can be observed directly in fig.2. The asymmetry which is
evident in fig.2a decreases as the mass cut is tightened.
When no mass cut is imposed the mass distribution is so deeply altered that
it is almost meaningless to compare the tree--level two lowest order moments
with those at NLO. This is still partially true for $\delta =  30$ GeV.
The low mass tail is non--negligible and considerably increase the mass
shift from .274 MeV to 1.70 GeV and the RMS width of the distribution,
from 2.39 GeV to 5.11 GeV. When the mass cut is  $\delta =  10$ GeV
$\Delta M$ increases from 41 MeV
to 125 MeV while the RMS deviation increase from about 1.7 GeV to slightly
more than 2 GeV.
\par
We believe that the impact of the results presented in this paper
on the precision measurement of $M_W$, both at threshold and through
direct reconstruction, should be carefully assessed taking into account
the other known sources of large corrections,
namely ISR and Coulomb interaction, and the experimental uncertainties.
In our opinion an improvement
of our understanding of the four--jet QCD background and a refinement of the
strategies to reduce it are required .
As for many other effects, if the theoretical analysis
can be made sufficiently precise and, at the same time, accurate data are
available, through a careful extrapolation of LEP I results,
or can be obtained, possibly through a dedicated run below
threshold, QCD backgrounds could become part of the predictions
without compromising the accuracy with which the $W$--mass will be measured.
Further work will be necessary to determine which combination of cuts
will result in the highest accuracy on $M_W$, balancing the smaller numer
of events with a decreased level of background.

\subsection*{Conclusions}
We have presented the calculation of the complete $\cal{O}(\alpha_s)$
corrections to $e^+e^- \ra WW \ra q_1 \bar q_1 q_2 \bar q_2$
in a form which allows to impose
realistic cuts on the structure of the observed events.
It has been shown that QCD radiation modifies in an important way
the distribution of jet--jet invariant masses and therefore that it strongly
influences the measurement of the $W$ mass from direct reconstruction of
the decay products.
For total hadronic cross--sections virtual and real contributions
cancel to a large degree if all events are retained. If however only events
with two jet--jet masses close to $M_W$ are accepted, higher order QCD
corrections can substantially decrease the cross section.\par
A number of issues have not been considered in this paper and will have to be
addressed in order to obtain a more complete theoretical understanding of the
production of $WW$ pairs at Lep II and to determine the best strategies
for the precision measurement of the $W$ mass.
(1) The precise choice
of the distribution or combination of variables from which to extract $M_W$.
(2) The influence of the jet reconstruction algorithm including hadronization.
(3) Initial state radiation. (4) Coulomb corrections. (5) Non resonant
background effects. (6) Colour reconnection effects.
As already mentioned, initial state radiation and Coulomb corrections
are relatively straightforward and an improved analysis which include both
effects is well under way.


\newpage

\subsection*{Figure Captions}

\begin{description}

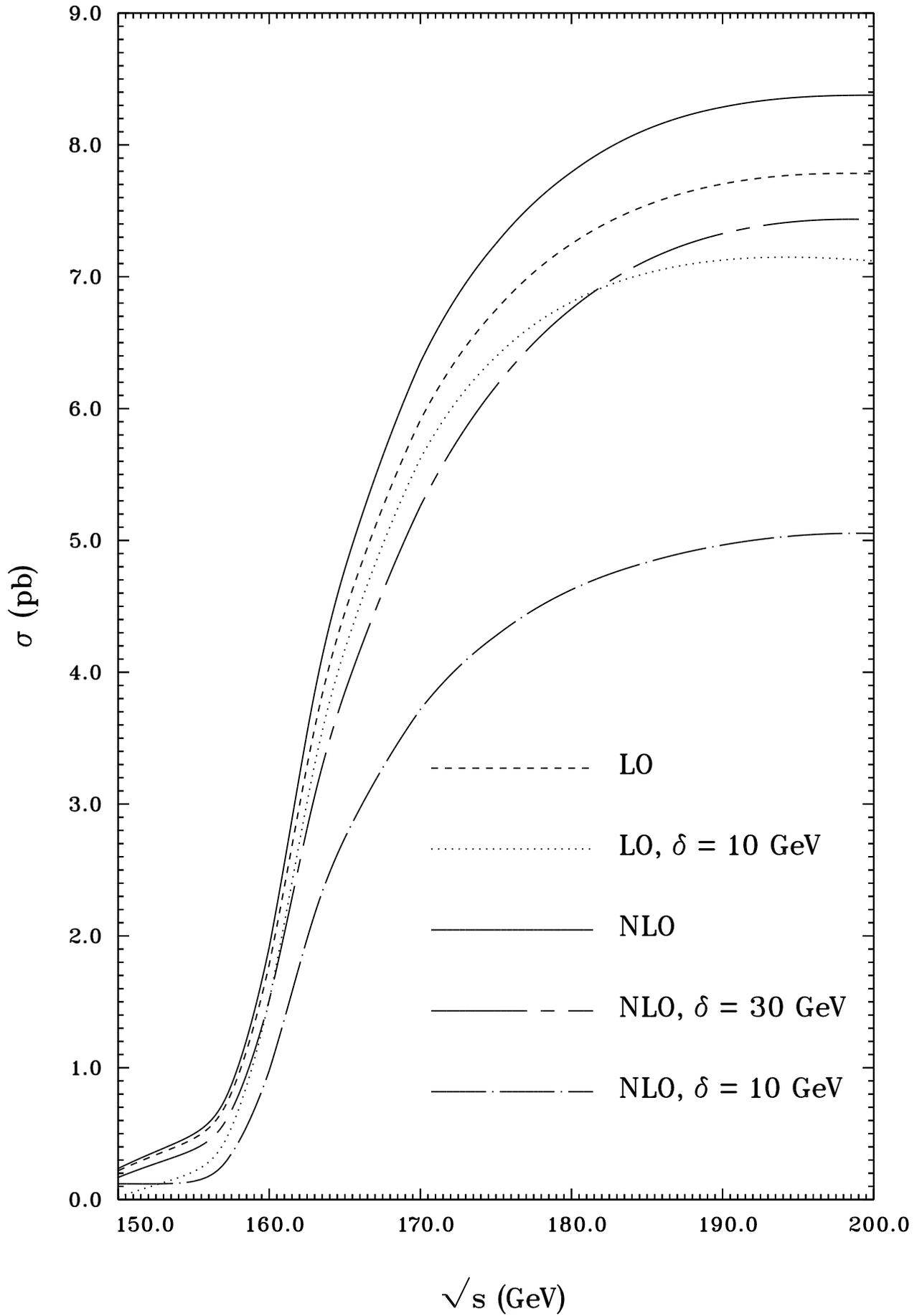
\item[Fig.1] Hadronic cross section for $e^+e^-\rightarrow W^+W^-$ as a
function of $\sqrt{s}$. The dashed line and the dotted line are
tree--level predictions. For the latter we have required
$ \left| M_{Ri} -M_W \right| \le 10 \ {\rm GeV},\ i=1,2 $.
The continuous line is the total hadronic NLO result.
The long--dash--short--dash curve and the dash--dot curve
show the NLO cross sections obtained with $\delta = 30$ GeV and
$\delta = 10$ GeV respectively.
ISR and Coulomb effects are not included.
The parameter values are given in Table II.

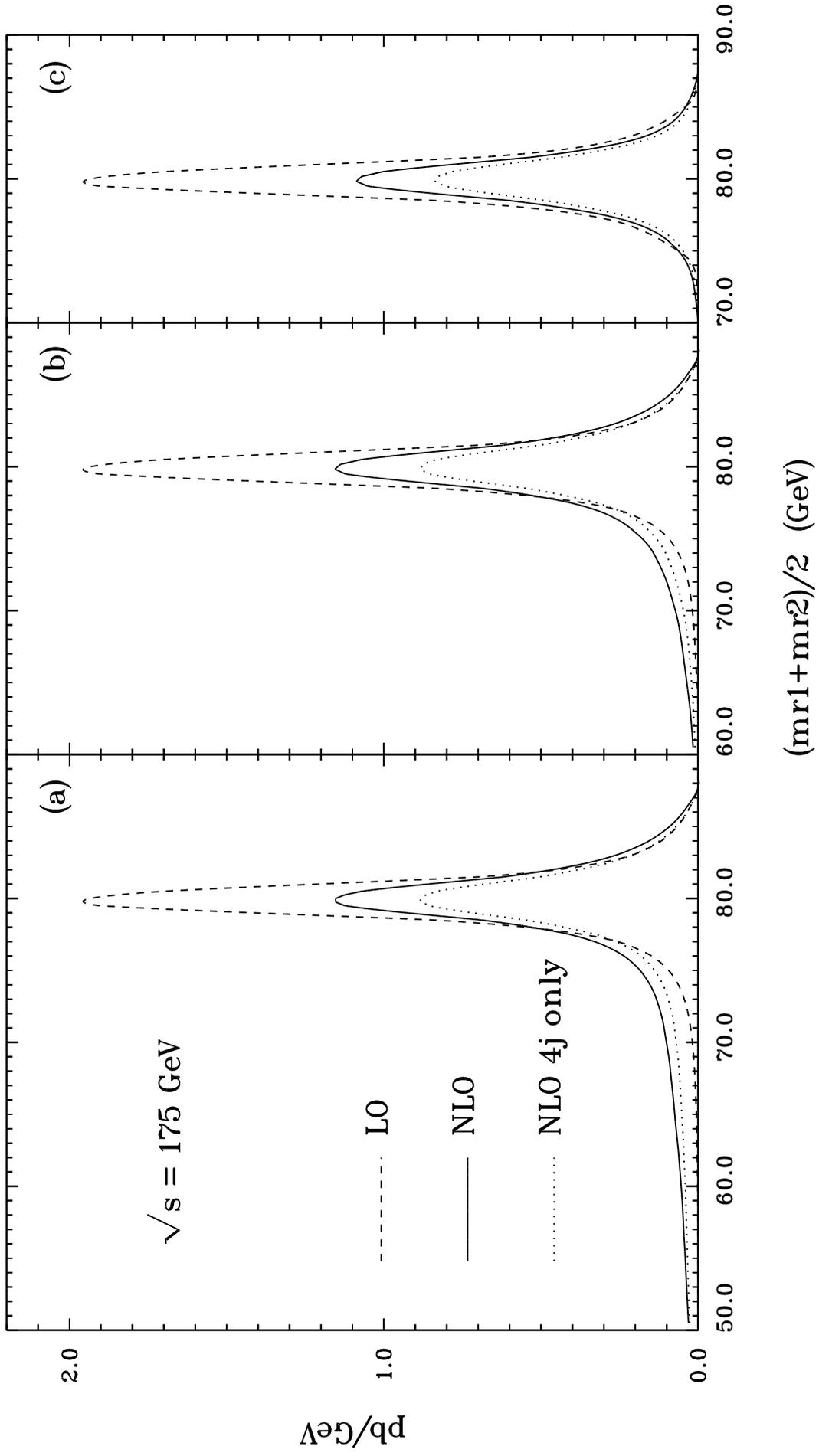
\item[Fig.2]
Distribution of the average of the two reconstructed mass
$\overline M = \frac{1}{2} \left( M_{R1} + M_{R2} \right)$.
The dashed lines are the tree--level predictions.
The continuous lines and the dotted lines include NLO QCD corrections.
For the latter five--jet events, defined using
$y_{cut} = 1. \times 10^{-2}$ in the
JADE scheme, have been excluded. In fig.2a no cut on
$ \left| M_{Ri} -M_W \right|$ has been imposed. In fig.2b and fig.2c
we have required
$ \left| M_{Ri} -M_W \right| \le 10$ GeV $i=1,2$
with $\delta = 30$ GeV and $\delta = 10$ GeV respectively.
ISR and Coulomb effects are not included.
The parameter values are given in Table II.

\end{description}

\vfill

\subsection*{Table Captions}

\begin{description}

\item[Table I] Branching ratios of the $WW$ pairs at tree--level. The
decay channels involving $\tau$'s are particularly challenging and are
usually analized separately.

\item[Table II] Parameters used in the numerical part of the paper.

\item[Table III] Total hadronic cross sections. For each energy the upper line
refers to the LO and the lower one to the results at NLO in QCD.
In the first column all events are accepted while in the second and third
the two jet--jet masses satisfy
$ \left| M_{Ri} -M_W \right| \le \delta, i=1,2 $ with
$\delta = 30$ GeV and $\delta = 10$ GeV respectively. The numbers in
parentheses are obtained discarding all events which are recognized as
five jets with $y_{cut} = 1.\times 10^{-2}$ in the JADE scheme.

\item[Table IV] Average deviation and RMS deviation from $M_W$ in GeV.
In each column the left hand side value refers to the LO
and the right hand side
one to the results at NLO in QCD.
In the first row all events are accepted while in the second and third
row the two jet--jet masses satisfy
$ \left| M_{Ri} -M_W \right| \le \delta, i=1,2 $ with
$\delta = 30$ GeV and $\delta = 10$ GeV respectively. The results
in the last three rows are obtained discarding
all events which are recognized as
five jets with $y_{cut} = 1.\times 10^{-2}$ in the JADE scheme.

\end{description}

\newpage

{
\renewcommand {\arraystretch}{1.5}

\begin{center}
\begin{tabular}{|c|c|}
\hline
$W^+ W^- \ra q_1 \bar q_1 q_2 \bar q_2$ &
                     $\frac{4}{9}$ \\ \hline
$W^+ W^- \ra q \bar {q} \ell \nu $ &
                     $\frac{8}{27}$ \\ \hline
$W^+ W^- \ra \ell \nu \ell^\prime \nu^\prime $ &
                     $\frac{4}{81}$ \\ \hline
$W^+ W^- \ra \tau + {\rm anything}$ &
                     $\frac{17}{81}$ \\ \hline
\end{tabular}
\end{center}
\begin{center}
Table I
\end{center}

\vspace{1. in}

\begin{center}
\begin{tabular}{|c|c|}
\hline
parameter & value \\
\hline \hline
$M_Z$ & 91.173 GeV \\ \hline
$M_W = M_Z \cos\theta_W$ & 79.905 GeV \\ \hline
$\Gamma_Z$ & 2.49 GeV \\ \hline
$\Gamma_W$ & 2.08 GeV \\ \hline
$\alpha^{-1}$ & 128. \\ \hline
$\sin^2\theta_W$ & 0.2319 \\ \hline
$\alpha_s$ & .117 \\ \hline
$(\hbar c)^2$ & .38937966 $10^9$ $pb$GeV$^2$ \\ \hline
\end{tabular}
\end{center}
\begin{center}
Table II
\end{center}

\vspace{1. in}

\begin{center}
\begin{tabular}{|c|c|c|c|}
\hline
 &  & $\delta = 30$ GeV & $\delta = 10$ GeV \\ \hline\hline
{$\sqrt{s} = 175 $ GeV} &
          6.749 & 6.725 & 6.393 \\
\cline{2-4}
\hspace{1cm}$(y_{cut} = 10^{-2})$ &
                7.251 (5.638) & 6.169 (4.718) & 4.277 (3.383)\\ \hline
{$\sqrt{s} = 160 $ GeV} &
          1.784 & 1.757 & 1.513 \\
\cline{2-4}
        & 1.916 & 1.513 & 0.980 \\ \hline
\end{tabular}
\end{center}
\begin{center}
Table III
\end{center}

\newpage

\begin{center}
\begin{tabular}{|c|c c|c c|}
\hline
$\sqrt{s} = 175 $ GeV &\multicolumn{2}{|c|}{$\langle \overline M - M_W\rangle$}
&\multicolumn{2}{|c|}
{$\langle\left(\overline M - M_W\right)^2\rangle^{\frac{1}{2}}$}  \\
\hline
 & LO & NLO & LO & NLO \\ \hline
           no cut                   & -0.340 & -4.82  & 2.645 & 10.65 \\ \hline
         $\delta = 30$ GeV          & -0.274 & -1.70  & 2.390 & 5.11 \\ \hline
         $\delta = 10$ GeV          & -0.041 & -0.125 & 1.679 & 2.02 \\ \hline
          $y_{cut} = 0.01$          & -0.340 & -3.88 & 2.645 & 9.88 \\ \hline
$\delta = 30$ GeV, $y_{cut} = 0.01$ & -0.274 & -1.09 & 2.390 & 4.15 \\ \hline
$\delta = 10$ GeV, $y_{cut} = 0.01$ & -0.041 & -0.092 & 1.679 & 1.78 \\ \hline
\end{tabular}
\end{center}
\begin{center}
Table IV
\end{center}

}

\end{document}